\documentclass[amssymb,prb,twocolumn,showpacs]{revtex4}
\usepackage{amsmath}
\usepackage{graphicx}
\usepackage{verbatim}
\usepackage{graphics}
\usepackage [T1] {fontenc}
\begin{document}

\title{Spin-polarized currents in double and triple quantum dots driven by ac magnetic fields}
\author{Maria Busl and Gloria Platero}
\affiliation{Instituto de Ciencia de Materiales de Madrid, CSIC,
Cantoblanco, 28049 Madrid,
Spain}

\begin{abstract}
\vspace{0.5cm} We analyze transport through both a
double quantum dot and a triple quantum dot with inhomogeneous Zeeman splittings 
in the presence of crossed dc and ac magnetic fields. We find that strongly spin-polarized current
can be achieved by tuning the relative energies of the Zeeman-split
levels of the dots, by means of electric gate voltages:
depending on the energy level detuning, the double quantum dot works either as
spin-up or spin-down filter. We show that a triple quantum
dot in series under crossed dc and ac magnetic fields can act not
only as spin-filter but also as spin-inverter. 
\end{abstract}
\maketitle

\section{Introduction}
A key aim in spintronics is the realization of spin-based quantum
information devices, where coherent electron spin manipulation is a
fundamental issue.\cite{loss,cota} 
In semiconductor quantum dots, coherent electron spin manipulation can
be realized by electron spin resonance (ESR), where an oscillating magnetic field
is applied to the sample in order to rotate the electron spin.\cite{koppens, loss1,rafa, levi, danon, busl}
Together with ESR, electron dipole spin
resonance techniques --- which combine ac electric fields with
spin-orbit interaction\cite{novak} or with a dc magnetic field
gradient\cite{pioro} --- have been implemented in order to measure
coherent rotations of one single electron spin\cite{koppens, novak}
in double quantum dots (DQDs). Coherent spin rotations of one single spin have also been 
proposed theoretically in triple quantum dots (TQDs)\cite{buslphyse} 
under crossed ac and dc magnetic fields. 

In ESR experiments in quantum dot arrays, an important issue is to individually address
the electron spin in each quantum dot. 
To this end, it has been proposed to tune the Zeeman splitting, in order to
manipulate the electron spin independently in each dot.\cite{coish}
The Zeeman splitting in a quantum dot is determined by
the intensity of the applied dc magnetic field and the electron g factor, 
$\Delta_{\text{Z}} = g\mu_{\text{B}}B_{\text{dc}}$. Hence different Zeeman splittings can occur
in quantum dot arrays where the dots have different g factors, or as well by applying different magnetic fields 
to each quantum dot. Both alternatives have been realized experimentally very recently: 
vertical DQDs made out of different materials --- e.g. GaAs and InGaAs --- 
show different g factors\cite{huang} and
on the other hand, in a sample with a spatially homogeneous g factor, 
an additional micro-ferromagnet placed nearby creates a different external magnetic 
field $B_{\text{dc}}$ in each dot.\cite{pioro}

The next logic step from DQDs to networks of quantum dots is a TQD, in linear or triangular arrangement.
Both versions have been realized experimentally in the last few years,\cite{ludwig,haug,gaudreau}
where tunneling spectroscopy and stability diagram measurements have been performed in order to
gain a deeper insight into the electronic configurations in TQDs, which is necessary for
potential three-spin qubit applications. On the theoretical side, next to fundamental studies of their
eigenenergy spectrum,\cite{hawrylak} TQDs have attracted interest mostly in a
triangular arrangement, where the system symmetry gives rise to fundamental coherence phenomena.
In this context, so-called ``dark states''\cite{emary,christina,busl} and Aharonov-Bohm oscillations\cite{busl,hawrylakAB}
have been studied. TQDs have been used
as a testing ground for Kondo physics\cite{kondo} and have been proposed as current rectifiers\cite{vidan,bulka} 
and spin-entanglers.\cite{saraga} 

In the present work, we are interested in single electron manipulation, and therefore study theoretically 
transport through both double and triple quantum dots. We calculate the current and 
current spin polarization through a DQD and a linear TQD array exposed to crossed dc and 
ac magnetic fields. We consider an inhomogeneous dc magnetic field
that produces different Zeeman splittings in the dots, while the g factor is the same in both dots.
For DQDs, a regime is considered where the 
system is occupied either by zero or one electron. For TQDs, the corresponding 
features are discussed for one or two electrons in the system.
With the single electron spin levels resolved in each quantum dot,
interdot tunneling is governed by definite spin selection rules, i.e. tunneling from one dot to the other
is only possible when two equal spin levels are aligned. However, when an ac magnetic field is applied,
it rotates the spin and allows for spin-flip processes along the tunneling that can lead to new features in
the current. This effect of an ac magnetic field has been explored previously in a DQD,\cite{koppens}
where the authors report single electron spin rotations by a combination of an ac magnetic field and sharp electric pulses.

In our work, as will be discussed in more detail below, we will focus our attention on the
polarizing effect of an ac field, i.e. we will show that the combination
of inhomogeneous dc and ac magnetic fields in DQDs and TQDs allows
for the creation of spin-polarized currents and thus for the design
of spin-filters and spin-inverters.

The paper is organized as follows: 
In section \ref{model} we introduce the model and the technique used to calculate transport
through a DQD and TQD. Section \ref{results} discusses in detail the results of this paper. 
In section \ref{undriven} we first briefly review 
the main result of a related experimental work that has recently been reported in the literature
and is important for further understanding. We then proceed in the following paragraphs \ref{omega1} - \ref{nonresonant}
with a detailed analysis of the main results of this paper, namely the spin-polarized currents 
produced by a combination of dc and ac magnetic fields at certain interdot level detunings.
The role of the system parameters involved in the polarization mechanism --- Zeeman splitting difference,
ac field amplitude and frequency and interdot tunneling amplitude --- is discussed. 
In section \ref{tqd} we present the corresponding results obtained for a TQD.
We end with a summary of the main results in section \ref{conclusions}.

\section{Model and technique}\label{model}
We consider a quantum dot array as shown schematically in
Fig.~\ref{schemainverter}.
The dots are coupled to each other
coherently by a tunneling amplitude $t_{\text{ij}}$ and are weakly
connected to source and drain contacts by rates $\Gamma_{\text{L}}$ and $\Gamma_{\text{R}}$.
The total Hamiltonian of the system is: 

\begin{figure}[t]
\begin{center}
\includegraphics[width=3in,clip] {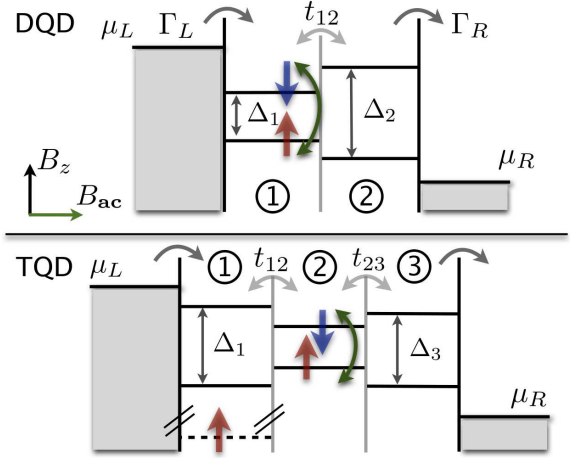}
\end{center}
\caption{\label{schemainverter}\small
(Color online) Schematic diagram of a DQD (above) and TQD (below) exposed to crossed dc 
($B_{\text{dc}}$) and ac ($B_{\text{ac}}$) magnetic fields.
The electron spin is rotated once the ac frequency matches the Zeeman
splitting in one of the dots. In the TQD, one electron is confined in the left dot (dot 1),
such that only an electron with opposite spin can enter the TQD. 
The dots are coupled coherently by tunneling amplitudes $t_{\text{ij}}$ and
incoherently to leads by rates $\Gamma_{\text{L}}$ and $\Gamma_{\text{R}}$.}
\end{figure}

\begin{equation}
\mathcal{H} = \mathcal{H}_{\text{Dots}}^0+\mathcal{H}_{t_{\text{ij}}
}^0+\mathcal{H}_{\text{B}}^0(t)+\mathcal{H}_{\text{T}}+\mathcal{H}_{\text{Leads}},
\end{equation}

where the individual terms are
\begin{align}
\mathcal{H}_{\text{Dots}}^0 &= \sum_{i\sigma}\xi_{i\sigma}\hat{c}^{\dagger}_{i\sigma}
\hat{c}_{i\sigma}+\sum_{i}U_{i}\hat{n}_{i\uparrow}\hat{n}_{i\downarrow}
+\frac{1}{2}\sum_{i\neq j}V\hat{n}_{i}\hat{n}_{j}\nonumber\\
\mathcal{H}_{t_{\text{ij}}}^0 &= -\sum_{ij\sigma}t_{\text{ij}}(\hat{c}^{\dagger}_{i\sigma}
\hat{c}_{j\sigma}+\hat{c}^{\dagger}_{j\sigma}\hat{c}_{i\sigma})\nonumber\\
\mathcal{H}_{\text{Leads}} &= \sum_{l\in{L,R},k\sigma}\epsilon_{lk\sigma}\hat{d}^{\dagger}_{lk\sigma}
\hat{d}_{lk\sigma}\\\nonumber
\mathcal{H}_{\text{T}} &= \sum_{l\in{L,R},k\sigma}\gamma_{l}(\hat{d}^{\dagger}_{lk\sigma}
\hat{c}_{l\sigma}+\hat{c}^{\dagger}_{l\sigma}\hat{d}_{lk\sigma}).\nonumber
\end{align}
The first term, $\mathcal{H}_{\text{Dots}}^0$, describes an isolated array of quantum dots, with electrons
coupled electrostatically. Here, $\xi_i$ stands for the single energy spectrum of an electron located in
dot $i$, and $U_i$ and $V$ are are the intra- and the inter-dot Coulomb repulsion respectively. 
$\mathcal{H}_{t_{\text{ij}}}^0$ describes the coherent tunneling between the dots, which in the case of a DQD
is given by $t_{12}$ and in a TQD by $t_{12}$ and $t_{23}$.
The quantum dot array is coupled to leads which are described by $\mathcal{H}_{\text{Leads}}$, and the 
coupling of the array to the leads is given by $\mathcal{H}_{\text{T}}$.
The magnetic field Hamiltonian consists of two parts, coming from a dc field $B_{\text{dc}}$ in $z$-direction, and an ac field
$B_{\text{ac}}$  applied in $xy$-direction:
\begin{equation}
\mathcal{H}_{\text{B}}^0(t)=\sum_{i}\left[\Delta_iS_{zi}+B_{\text{ac}}(\cos(\omega t)S_{xi}+ \sin(\omega t)S_{yi})\right],
\end{equation} being ${\bf
S}_{i}=\frac{1}{2}\sum_{\sigma\sigma'}\hat{c}^{\dagger}_{i\sigma}
\sigma_{\sigma\sigma'}\hat{c}_{i\sigma'}$ the spin operator of the
$i$th dot, and the sum running over index $i = 1,2$ for the DQD and $i=1,2,3$ for the TQD. 
$B_{\text{dc}}$ has a different intensity in each dot and
thus produces different Zeeman splittings $\Delta_{i} = g\mu B_{\text{dc}_i}$, while we
consider the dots with equal g factor. $B_{\text{ac}}$ induces
spin rotations when its frequency fulfills the resonance condition
$\omega = \Delta_{i}$. The time-dependent Hamiltonian can be
transformed by means of a unitary transformation\cite{rafa,busl}
$U(t) =\text{exp}(-i[\omega t\sum_{i}S_{zi}])$ into the rotating
reference frame. The resulting time-independent Hamiltonian is then:
\begin{equation}
\mathcal{H}
_{\text{B}}^0=\sum_{i}\left[(\Delta_i-\hbar\omega)S_{zi}+B_{\text{ac}}S_{xi}\right].
\end{equation}
The dynamics of the system is given by the time evolution of the
reduced density matrix elements $\rho_{mn}$, whose equations of
motion read, within the Born-Markov-approximation:
\begin{align}
\dot{\rho}_{mn}(t)=&-i\langle m|[\mathcal{H}
_{\text{Dots}}^0+\mathcal{H}_{t_{\text{ij}}}^0+\mathcal{H}_{\text{B}}^0,\rho]|n\rangle\\
&+\sum_{k\neq{n}}(\Gamma_{nk}\rho_{kk}-\Gamma_{kn}\rho_{nn})\delta_{mn}\nonumber\\
&-\Lambda_{mn}\rho_{mn}(1-\delta_{mn})\nonumber
\label{density}
\end{align}
The commutator accounts for the coherent dynamics in the quantum dot array,
tunneling to and from the leads is governed by transition rates $\Gamma_{mn}$ 
from state $|n\rangle$ to state $|m\rangle$, and 
decoherence due to interaction with the reservoir is considered in the term 
$\Lambda_{mn}=\frac{1}{2}\sum_{k}
(\Gamma_{km}+\Gamma_{kn})$. 
The transition rates are calculated using Fermi's golden rule:
\begin{align}
\Gamma_{mn} = \sum_{l=L,R}\Gamma_{l}&[f(E_m-E_n-\mu_l)\delta_{N_{m},N_{n}+1} +\nonumber\\
&(1-f(E_m - E_n - \mu_l))\delta_{N_m,N_{n}-1}],
\end{align}
where $E_{m}-E_{n}$ is the energy difference between states $|m\rangle$ and $|n\rangle$ of the isolated 
quantum dot array, and $\Gamma_{\text{L,R}} = 2\pi\mathcal{D}_{\text{L,R}}\vert\gamma_{\text{L,R}}\vert^2$ are the tunneling rates for each lead. 
The density of states $\mathcal{D}_{\text{L,R}}$ and the tunneling couplings $\gamma_{\text{L,R}}$ are assumed to be 
energy independent. We set $\Gamma_{\text{L}} = \Gamma_{\text{R}} = \Gamma$. 

We consider strong Coulomb repulsion, such that the DQD can be occupied with at most one extra
electron. It is then described by a basis of 5 states, namely:
$|0,0\rangle$, $|{\uparrow},0\rangle$, $|{\downarrow},0\rangle$,
$|0,{\uparrow}\rangle$, $|0,{\downarrow}\rangle$. With a bias
applied from left to right, current $I$ flows whenever dot 2 is
occupied: 
\begin{equation}
I_{\text{DQD}} = \Gamma(\rho_{|0,{\uparrow}\rangle} +\rho_{|0,{\downarrow}\rangle})
\end{equation}
In the TQD, one electron is confined in the left dot (dot 1, see Fig.~\ref{schemainverter}, lower panel), 
and the chemical potential of the left lead is such that
only an electron with the opposite spin can enter the TQD. Considering here as well strong Coulomb
repulsion, we allow only for one additional electron to enter the TQD. 
The full two-electron basis for the TQD contains fifteen two-electron states, and one zero- and
six one-electron states. For the scope of this paper, it is sufficient to look at transport around 
the triple point $(2,0,0)\leftrightarrow(1,1,0)\leftrightarrow(1,0,1)$. 
The number of relevant basis states is then reduced to eleven, which are
\begin{itemize}
\item 1-electron states: $|{\uparrow},0,0\rangle$, $|{\downarrow},0,0\rangle$
\item 2-electron states: $|{\uparrow},{\uparrow},0\rangle$, $|{\uparrow},{\downarrow},0\rangle$, $|{\downarrow},{\uparrow},0\rangle$, $|{\downarrow},{\downarrow},0\rangle$\\
$|{\uparrow},0,{\uparrow}\rangle$, $|{\uparrow},0,{\downarrow}\rangle$, $|{\downarrow},0,{\uparrow}\rangle$, $|{\downarrow},0,{\downarrow}\rangle$\\
$|{\uparrow}{\downarrow},0,0\rangle$.
\end{itemize}
The current from left to right through the TQD is calculated summing over all states that include an electron in the right dot (dot 3):
\begin{equation}
I_{\text{TQD}} = \Gamma\left(\rho_{|{\uparrow},0,{\uparrow}\rangle} + \rho_{|{\uparrow},0,{\downarrow}\rangle} + \rho_{|{\downarrow},0,{\uparrow}\rangle} + \rho_{|{\downarrow},0,{\downarrow}\rangle}\right)
\end{equation} 
The spin-resolved currents hence are
\begin{align}
I_{\uparrow} &= \Gamma\left(\rho_{|{\uparrow},0,{\uparrow}\rangle} + \rho_{|{\downarrow},0,{\uparrow}\rangle} \right)\nonumber\\
I_{\downarrow} &= \Gamma\left(\rho_{|{\uparrow},0,{\downarrow}\rangle} + \rho_{|{\downarrow},0,{\downarrow}\rangle} \right).
\end{align} 

The spin polarization of the current is defined as 
\begin{equation}
P = \frac{I_{\uparrow} - I_{\downarrow}}{I_{\uparrow} + I_{\downarrow}},
\end{equation} 
where $I_{\uparrow}(I_{\downarrow})$ is the $\uparrow$($\downarrow$)-current.

\section{Results}\label{results}
\subsection{Undriven case: $B_{\text{ac}} = 0$}\label{undriven}
Let us now start to describe transport through a DQD (see Fig.~\ref{schemainverter}, upper panel).
In this section we will reproduce within our theoretical framework the results 
recently reported by Huang {\it et al.}.\cite{huang} The authors have shown
that in transport through DQDs with different Zeeman splittings a so-called spin bottleneck
situation can occur: When either ${\uparrow}$- or ${\downarrow}$-levels
are aligned, transport is suppressed, whereas the current is largest
in the configuration where the interdot level detuning $\epsilon$ is set to half
the Zeeman energy difference.

\begin{figure}[t]
\begin{center}
\includegraphics[width=3in,clip]{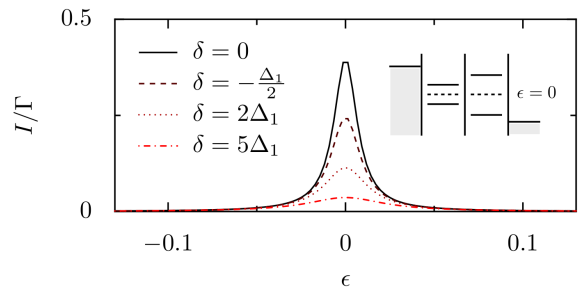}
\end{center}
\caption{\label{invfig2}\small (Color online) Current $I$ versus detuning $\epsilon$ in an
undriven DQD with different Zeeman splittings. Here maximal current
flows, when $\epsilon = 0$, and this central current decreases for increasing $\delta$,
since then parallel spin-levels are more separate.
Parameters ($e=\hbar=1$, in meV): $t_{\text{12}}
 = 0.005$,
$\Gamma = 0.001$, $\Delta_1 = 0.025$ ($B_{\text{dc}}\approx 1$T), and the
current $I$ is normalized in units of the hopping $\Gamma$ to the
leads.}
\end{figure}

Applying a dc magnetic field in z-direction produces a Zeeman
splitting $\Delta_z$, which we consider inhomogeneous:
$\Delta_1 \neq \Delta_2$, and $\delta = \Delta_2 - \Delta_1$. 
If an electron tunnels onto the ${\uparrow}$(${\downarrow}$)-level in dot 1 
that is far from resonance from the corresponding spin-level in dot 2, a spin blockade or
bottleneck situation arises: Spin is conserved at tunneling, so the
electron remains in dot 1 without being able to tunnel to dot 2.
This {\it blockade} is only relieved by a finite level broadening
and coupling to the leads. The maximal current
occurs then for the most symmetric level arrangement, that is when
neither ${\uparrow}$- nor ${\downarrow}$-levels are in resonance, but
when they are symmetrically placed around each other (see Fig.~\ref{invfig2}).
Increasing the Zeeman splitting difference $\delta$ 
maintains the bottleneck situation, but the central current decreases, since 
it is a consequence of the level hybridization of the same
spin-levels due to tunneling. Hence, the further separated they are,
the less current flows. Notice that the current only depends on the
Zeeman splitting difference $\delta$ and not on the absolute values.

Interdot tunneling conserves spin and the current through the sample is
completely unpolarized. In ac magnetic fields however, the electron
spin undergoes rotations and the spin selection rules thus do not apply any more. 
For certain detunings, this will lead to spin-polarized currents, as we
will see in the next section.

\subsection{Resonance condition: $\omega = \Delta_1$}\label{omega1}
With a circularly polarized ac magnetic field $B_{\text{ac}}$ applied to the DQD, the
transformed Hamiltonian $\mathcal{H}^0$ reads:
\begin{equation}
\mathcal{H}^0=\left(
\begin{array}{cccc}
-\frac{\Delta_1}{2} + \frac{\omega}{2} & \frac{B_{\text{ac}}}{2} & -t_{\text{12}}
 & 0 \\
 \frac{B_{\text{ac}}}{2} &  \frac{\Delta_1}{2} - \frac{\omega}{2} & 0 & -t_{\text{12}}
 \\
 -t_{\text{12}}
 & 0 & -\frac{\Delta_2}{2} + \frac{\omega}{2} - \epsilon & \frac{B_{\text{ac}}}{2}  \\
0 & -t_{\text{12}}
 &  \frac{B_{\text{ac}}}{2} & \frac{\Delta_2}{2} - \frac{\omega}{2} - \epsilon
\end{array}
\right),
\label{hamildqd}
\end{equation}
where $\epsilon$ is the detuning between dot 1 and dot 2.

For the ease of its analysis, Hamiltonian \eqref{hamildqd}
can be seen as a a pair of two-level systems coupled by $t_{12}$. 
In a two-level system, the important physical
quantities are the energy difference (``detuning'') of the two
levels and the coupling between them. In the present case, note that $t_{\text{12}}$ 
couples only levels with the same spin,
which are detuned by $\pm\delta/2 + \epsilon$, where $\delta
= \Delta_2 - \Delta_1$. Moreover, within each dot the
different spin-levels are coupled by $B_{\text{ac}}/2$ and
``detuned'' by $\omega - \Delta_{1,2}$ (see diagonal elements in
\eqref{hamildqd}). Therefore, depending on the ac frequency $\omega$,
the energy levels in either left or right dot are renormalized to the same energy. In the other dot however,
since there $\omega \gtrless \Delta_{i}$, the renormalized splitting between the spin-levels 
becomes smaller when $\omega<\Delta_{i}$ or bigger for $\omega>\Delta_i$.
We will focus first on the resonance condition $\omega = \Delta_1$, as it is the most relevant here.

\begin{figure}[t]
\begin{center}
\includegraphics[width=3in,clip]{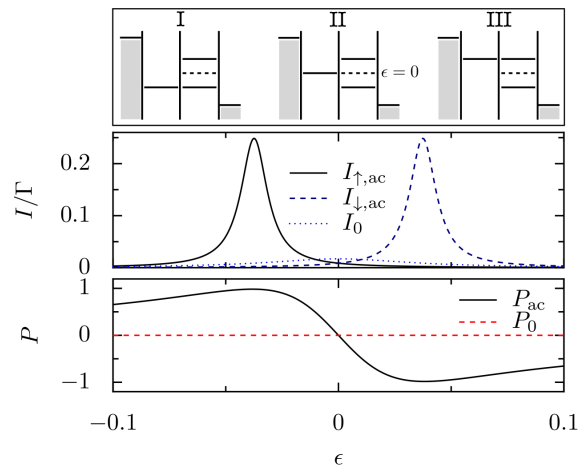}
\end{center}
\caption{\label{invfig3}\small (Color online) Upper panel:
Energy level distribution for different detunings $\epsilon$ in a DQD driven by
$B_{\text{ac}}$. When $\omega = \Delta_1$, the levels in dot 1
renormalize to the same energy (their eigenenergies are split by $B_{\text{ac}}$, see
text), and the levels in dot 2 get closer or farther apart than in
the undriven case.
Middle panel: Spin-resolved currents $I_{\uparrow}$ and $I_{\downarrow}$
vs. detuning $\epsilon$. At $\epsilon\approx \pm\delta/2$, the current is
strongly $\uparrow$($\downarrow$)-polarized, compared to the undriven current $I_0$. 
Lower panel: Polarization $P$ versus the detuning $\epsilon$. Note the strong
polarization ($P\approx\pm1$) around $\epsilon\approx
\pm\delta/2$. 
Parameters in meV ($e = \hbar = 1$): 
$\Gamma = 0.001$, $t_{\text{12}} = 0.005$, $B_{\text{ac}} = 0.005$ ($\approx 0.2$T), 
$\Delta_1 = 0.025$ ($B_{\text{z}1}\approx 1$T), $\Delta_2 = 0.1$.}
\end{figure}

In order to understand the effect of $B_{\text{ac}}$ on the system,
let us look at the eigenstates of the isolated dots 1 and 2. In dot
1, since $\omega = \Delta_1$, the eigenstates are
$|\psi_1\rangle^{\pm} = \frac{1}{\sqrt{2}}( |{\uparrow}_1\rangle \pm
|{\downarrow}_1\rangle)$ and their eigenenergies differ by
$B_{\text{ac}}$. In dot 2 however, since it is out of resonance, the
eigenstates depend both on $\delta$ and $B_{\text{ac}}$:
\begin{equation}
|\psi_2\rangle^{\pm} = \frac{1}{N^{\pm}}(-\frac{\delta\pm\sqrt{B_{\text{ac}}^2+\delta^2}}{B_{\text{ac}}}|\uparrow_2\rangle +|\downarrow_2\rangle)
\label{states}
\end{equation}
Here $N^{\pm} =
\sqrt{2}/B_{\text{ac}}\left(\sqrt{B_{\text{ac}}^2+\delta^2\pm\delta\sqrt{B_{\text{ac}}^2+\delta^2}}\right)$
are the normalization factors. The eigenenergies associated to these states are separated
by $\sqrt{B_{\text{ac}}^2 + \delta^2}$. It is straightforward to
show that for $B_{\text{ac}} \ll\delta$, the eigenstates in dot 2
are almost pure ${\uparrow}$(${\downarrow}$)-states, i.e. the
spin-mixing is weak. Regarding the detuning $\epsilon$, we
distinguish three different level arrangements, see Fig.~\ref{invfig3}, upper panel:
In case I, the $\uparrow$- and
$\downarrow$-levels in dot 1 are aligned with the $\uparrow$-level
in dot 2, case II is the {\it symmetric} situation, and in case III
the levels in dot 1 are in resonance with the $\downarrow$-level in
dot 2.

In Fig.~\ref{invfig3}, lower panels, we plot the current $I$ through the driven
DQD and the polarization $P$ as a function of the level detuning $\epsilon$. It shows two
peaks at $\epsilon \approx \pm\delta/2$. At these lateral
peaks, corresponding to case I and III, the current is strongly {\it
spin-polarized}: an electron in dot 1 is rotated by the ac field which breaks
the spin bottleneck and the electron can thus tunnel to dot 2,
where the spin-levels are almost pure, or --- speaking in terms of the rotating field ---
the ac frequency in dot 2 is far off resonance and cannot rotate the electron there.
We thus arrive at one of the main result of this paper: under the condition $\omega = \Delta_1$,
dot 2 acts as a spin-filter, and it depends on $\epsilon$, whether
it filters $\uparrow$- or $\downarrow$-electrons. Notice that the current $I$
only depends on $\delta$ and not on the absolute values
$\Delta_{1,2}$.

For the purpose of a spin-filter, one has to answer the question as to how reliable the mechanism is,
and how it depends on the different system parameters. Both strong polarization and measurable 
currents are desirable. Here, we discuss the sensibility of the spin filtering mechanism towards the
interplay between tunneling $t_{\text{12}}$, ac field intensity
$B_{\text{ac}}$ and Zeeman splitting difference $\delta$. 

In order to get more insight into the problem, we obtain the current $I$ analytically for
certain limits: At symmetric detuning $\epsilon = 0$ (case II), the
current is unpolarized and reads
\begin{equation}
\frac{I_0}{\Gamma} = \frac{4t_{\text{12}}
^2 (4 B_{\text{ac}}^2+\Gamma^2+\delta^2)}{4 B_{\text{ac}}^2 (\Gamma^2+10 t_{\text{12}}
^2)+(\Gamma^2+\delta^2)(\Gamma^2+10 t_{\text{12}}
^2+\delta^2)}.
\label{stromcent}
\end{equation}
$I_0$ decreases for large $\delta$ and increases with growing
$B_{\text{ac}}$. In the limit of very large $t_{\text{12}}$, the
total current $I$ saturates to
$I/\Gamma(t_{\text{12}}\to\infty) = 2/5$. For the
limiting cases of $B_{\text{ac}}$ we get
\begin{align}
\lim_{B_{\text{ac}} \to0}\frac{I_{\uparrow,\downarrow}}{\Gamma} &= \frac{2t_{\text{12}}
^2}{\Gamma^2+10t_{\text{12}}
^2+4\epsilon^2+\delta^2}\label{limitb0}\\
\lim_{B_{\text{ac}} \to\infty}\frac{I_{\uparrow,\downarrow}}{\Gamma} &= \frac{2t_{\text{12}}
^2}{\Gamma^2+10t_{\text{12}}
^2+4\epsilon^2}\label{limitb1}.
\end{align}

For $B_{\text{ac}} \to 0$, i.e. in the undriven case, the current is
unpolarized and maximal at $\epsilon = 0$ and decreases for growing
$\delta$, see Eq.\eqref{limitb0}. Notice that in the opposite limit,
i.e. for large $B_{\text{ac}}$ (Eq.\eqref{limitb1}), the current is the same as in the
undriven case for $\delta = 0$. In this case, the difference of the eigenenergies
in each isolated dot becomes $B_{\text{ac}}$ in {\it both} dots and the
spins are mixed almost equally strongly. The polarized side-peaks therefore
disappear in favor of the unpolarized central current peak, see also
Eq.\eqref{stromcent}.  

Numerical analysis for intermediate field and tunneling amplitude yields that 
when $t_{\text{12}}$ and $B_{\text{ac}}$
become of the order of $\delta$, the current is practically unpolarized.
We find that at $B_{\text{ac}}/t_{12}\approx 1.5$, 
the polarization is strongest, when $\delta/t_{12}$
is at least one order of magnitude bigger than $B_{\text{ac}}/t_{12}$. 
It can be shown numerically that for $t_{\text{12}},B_{\text{ac}}\ll\delta$ 
the position $\epsilon$ of the side-peaks is
$\epsilon \approx \pm \delta/2$. The larger $\delta$, the
further separated the peaks corresponding to $I_{\uparrow}$ and
$I_{\downarrow}$. As a consequence, also the polarization is stronger for large
$\delta$, since the overlap of the spin-resolved currents tends to
zero. 

\begin{figure}[t]
\begin{center}
\includegraphics[width=3in,clip]{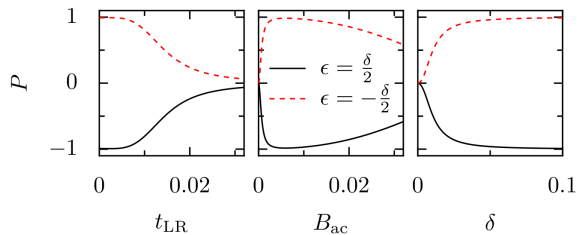}
\end{center}
\caption{\label{dqdbac3}\small (Color online) Polarization $P$ versus $t_{\text{12}}$ ,
$B_{\text{ac}}$ and $\delta$ in ac-driven DQD at detuning $\epsilon
= \pm\delta/2$ for $\omega = \Delta_1$: Left and middle
panel: For both small $t_{\text{12}}$ and $B_{\text{ac}}$,
spin-polarized current flows. $\vert P \vert$ becomes smaller as
$B_{\text{ac}}$ and $t_{\text{12}}$ grow. Right panel: $P$ is zero
at $\delta = 0$ and increases with $\delta$. Parameters see
Fig.~\ref{invfig3}.}
\end{figure}

In order to illustrate the effect of tunneling $t_{\text{12}}$,
ac field intensity $B_{\text{ac}}$ and Zeeman splitting difference $\delta$ on the polarization $P$, 
we calculate $P$ at $\epsilon = \pm\delta/2$ (Fig.~\ref{dqdbac3}). In the left and middle
panel, one can appreciate that for both small $t_{\text{12}}$ and
$B_{\text{ac}}$, $P\approx\pm1$, and it becomes smaller as
$t_{\text{12}}$ and $B_{\text{ac}}$ increase (for constant $\delta$). 
The right panel in Fig.~\ref{dqdbac3} shows the polarization for increasing $\delta$: the larger
$\delta$, the stronger $P$ .

\subsection{Resonance condition: $\omega = \Delta_2$}\label{omega2}
When the ac field instead fulfills the resonance condition $\omega = \Delta_2$, 
the energy renormalization due to $\omega$ is reversed in
the two dots as compared to $\omega = \Delta_1$, and now the energy
levels in dot 2 become degenerate. The analytical limits described for
$\omega = \Delta_1$ hold here as well: For large $t_{\text{12}}$ and
$B_{\text{ac}}$, the current becomes unpolarized, and at $\epsilon = 0$, it
follows Eq.\eqref{stromcent}. However, out of these limits,
transport behavior here is very different from the case $\omega =
\Delta_1$: At detunings $\epsilon \approx \pm \delta/2$, spin bottleneck 
occurs similar as was shown in the undriven case: Since dot 1 is out of resonance, the ac field can
not rotate the electron there, hence tunneling to dot 2 is strongly
suppressed. The maximal (unpolarized) current then flows for
$\epsilon = 0$ and no side-peaks appear.

\begin{figure}[t]
\begin{center}
\includegraphics[width=3in,clip]{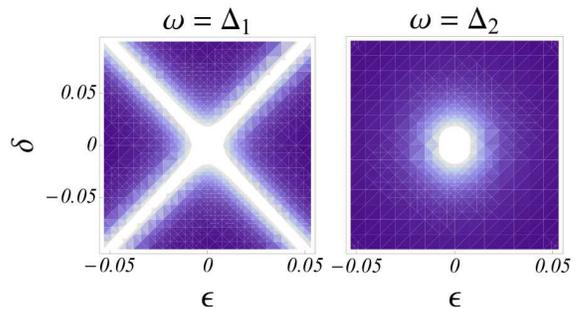}
\end{center}
\caption{\label{dqdbac4}\small (Color online) Density plots of the current $I$
versus detuning $\epsilon$ and Zeeman splitting difference $\delta$.
Left side: $\omega =\Delta_1$: For growing $\delta$ and $\epsilon$, the current $I$ splits off in
two branches (light-colored regions), which are spin-polarized in opposite direction (cf. previous section).
Right side: $\omega = \Delta_2$: Current flows only around $\delta = \epsilon = 0$ (light-colored region);
$P=0$. Parameters see Fig.~\ref{invfig3}. }
\end{figure}

In summary, at $\omega\approx\Delta_1$, dot 2 can always act as a
spin-filter. The mixing of $\uparrow$- and $\downarrow$-states due
to the ac field is always stronger in dot 1 than in dot 2, no matter
if $\Delta_1 \gtrless \Delta_2$. The ac field mixes $\uparrow$- and
$\downarrow$-states in dot 1 such that at $\epsilon \approx \pm \delta/2$, the
electron tunnels onto the almost pure $\uparrow$- or
$\downarrow$-levels in dot 2, which thus filters the spin and gives rise
to spin-polarized currents. This is opposed to the case $\omega = \Delta_2$: Here, due
to spin bottleneck, tunneling to dot 2 is only possible around
$\epsilon = 0$, where the current is totally unpolarized. This behavior
is shown in Fig.~\ref{dqdbac4} in two density plots of the current $I$ versus
detuning $\epsilon$ and $\delta = \Delta_2 - \Delta_1$ for the two cases
$\omega = \Delta_1$ (left) and $\omega = \Delta_2$ (right). In the
left plot, one can clearly see the formation of the two
spin-polarized current branches, which move far apart as $\delta$
and $\epsilon$ grow. In contrast to that, the right plot shows that current only flows for both
$\epsilon = 0$ and $\delta = 0$, and no spin-polarized side-peaks arise.

\subsection{Non-resonant driving}\label{nonresonant}
\begin{figure}[t]
\begin{center}
\includegraphics[width=3in,clip]{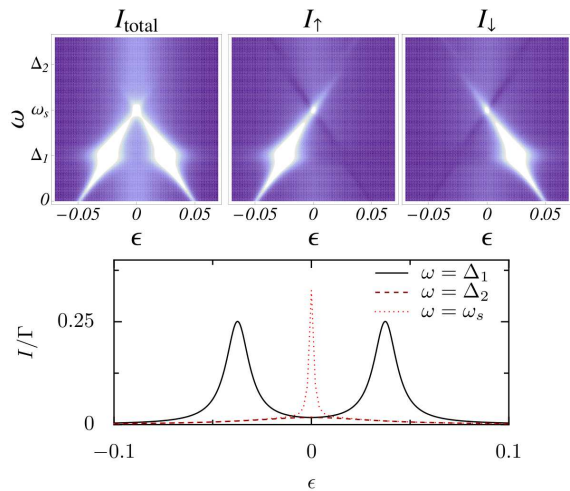}
\end{center}
\caption{\label{dqdbac6}\small (Color online) Upper panel: Density plots of the total current (left)
and spin-resolved currents $I_{\uparrow}$ (middle)
and $I_{\downarrow}$ (right) vs. detuning $\epsilon$ and ac frequency $\omega$ for $\Delta_1<\Delta_2$.
The lighter the color, the higher the current. 
Note that only very low current flows in the frequency range $\omega>\omega_{\text{s}}$ 
around $\epsilon = 0$. At $\omega = \omega_{\text{s}}$ and $\epsilon=0$, one sharp unpolarized peak arises. 
Lowering $\omega$ further, the current splits into two arms and successively grows,
until around $\omega = \Delta_1$, current is strongly enhanced and polarized, since the sidearms stem from
either $\uparrow$- or $\downarrow$-electrons, see middle and right upper panel.
Lower panel: Current versus $\epsilon$ for the three different situations $\omega = \Delta_1$, $\omega = \Delta_2$ 
and $\omega = \omega_{\text{s}}$. One can appreciate the big difference in the current intensities: Only for $\omega = \Delta_1$, 
polarized sidepeaks arise. For $\omega = \Delta_2$, current flows weakly around $\epsilon = 0$ and for $\omega_{\text{s}}$,
only at $\epsilon = 0$ a sharp current peak appears.
Parameters see Fig.~\ref{invfig3}.}
\end{figure}

If the ac frequency does not match
any of the Zeeman splittings $\Delta_{1,2}$, the effective finite
Zeeman splittings are $\Delta_{1,2}^* = \Delta_{1,2} - \omega$. It
is easy to prove that for $\omega = (\Delta_1 + \Delta_2)/2 =
\omega_{\text{s}}$, there is a ``symmetric'' situation, namely
$\Delta_{1}^* = (\Delta_1 - \Delta_2)/2$, and
$\Delta_{2}^*=-\Delta_{1}^*$. 
In this case, the mixing of the spin-states within each dot is equal in both
dots, or in other words, both dots are equally far from resonance
with the ac field. 
Regarding interdot tunneling, the levels are resonant
at $\epsilon = 0$, giving rise to one unpolarized current-peak. At
all other detunings $\epsilon$, spin bottleneck avoids the formation
of polarized side-peaks.
In Fig.~\ref{dqdbac6} we show the total current $I$ (upper left) and spin-resolved currents $I_{\uparrow}$ (upper middle), 
$I_{\downarrow}$ (upper right) vs. detuning $\epsilon$ and frequency
$\omega$, for $\Delta_1<\Delta_2$. In order to appreciate the different current intensities, we plot in the lower panel 
the total current versus the detuning $\epsilon$ for the three relevant frequencies $\omega = \Delta_1,\Delta_2,\omega_{\text{s}}$. 
Note the regimes for $\omega$, as discussed in the previous sections: 
For $\omega=\Delta_2$, spin bottleneck only allows for a very weak 
and unpolarized current to flow around $\epsilon = 0$. When the frequency matches the symmetric value $\omega_{\text{s}}$, at $\epsilon=0$
one sharp and unpolarized current peak arises, as predicted. 
Further decreasing of the frequency splits the current into two branches, 
which are enhanced and broadened as $\omega\approx\Delta_1$. 
The sidearms correspond to either $\uparrow$(middle panel)- or $\downarrow$(right panel)-electrons. 
For any off-resonant frequency, the current depends not only on $\delta$ as in the resonant case, 
but also on the absolute values $\Delta_{1,2}$. Hence the
position of the side-peaks is not
$\epsilon\approx\pm\delta/2$, but follows a different
behavior. This explains the kink in Fig.~\ref{dqdbac6} (upper panel) around
$\omega = \Delta_1$.

We want to stress that, in the ac-driven DQD, spin-polarized
currents can be achieved both for $\Delta_1 > \Delta_2$ or $\Delta_1
< \Delta_2$, since by varying the frequency $\omega$ one can always
tune one Zeeman splitting to be smaller than the other, as
schematically indicated by the renormalization of the energy levels
due to $\omega$ (see Fig.~\ref{invfig3}, upper panel). In contrast
to that, a static magnetic field set-up --- for example, considering
dc magnetic fields in $x$-direction\cite{tokura} ---  would only
produce polarized currents for $\Delta_1 < \Delta_2$.

\subsection{A triple quantum dot as spin-inverter}\label{tqd}
Now we want to implement the spintronic functionality of the spin-filter device
towards a {\it spin-inverter}, and to this end we consider a TQD. Our goal is to
produce spin-polarized incoming current $I_{\text{in}}$ and
oppositely spin-polarized outgoing current $I_{\text{out}}$. 

\begin{figure}[t]
\begin{center}
\includegraphics[width=3in,clip]{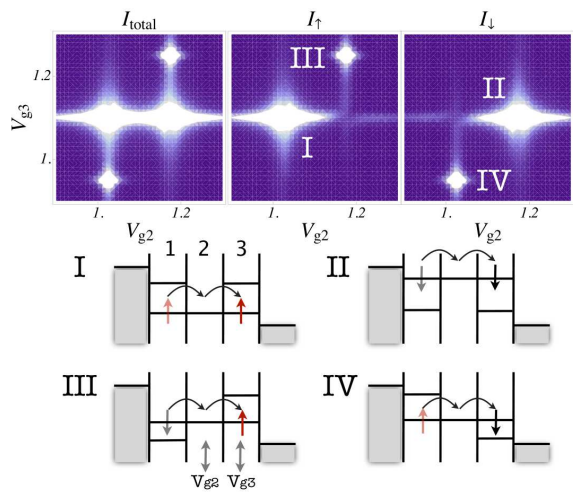}
\end{center}
\caption{\label{tqdinverter}\small (Color online) Total current $I$ and spin-resolved currents $I_{\uparrow}$ and $I_{\downarrow}$
vs. gate voltages $V_{\text{g2}}$ and $V_{\text{g3}}$ applied to the central dot (dot 2) and the right dot (dot 3) in a TQD
exposed to crossed $B_{\text{dc}}$ and $B_{\text{ac}}$. Here $\omega = \Delta_2$ and $\Delta_1 = \Delta_3 > \Delta_2$.
Four relevant level configurations can occur due to adjustment of $V_{\text{g2}}$ and  $V_{\text{g3}}$: 
in cases I and II, current through the TQD is polarized in one spin-direction,
and in cases III and IV, the electron spin is inverted. In order not to overload the figure, we indicate only the spins of the
incoming and outgoing electrons, but note that always one electron is confined in an off-resonant state in the left dot (dot 1, cf. Fig.~\ref{schemainverter}). 
Parameters in meV ($e = \hbar = 1$): $\Gamma = 0.01$, $t_{\text{12, 23}}= 0.01$, $B_{\text{ac}} = 0.01$ ($\approx 0.4$T), 
$\Delta_1 = \Delta_3 = 7\Delta_2$, $\Delta_2 = 0.025$ ($B_{\text{dc}} \approx 1$T), $U = 1.0$.}
\end{figure}

We consider the TQD in a regime where only 2 electrons can be in the TQD at a time,
and one electron is confined electrostatically in the left dot (dot 1, cf. Fig.~\ref{schemainverter}, lower panel). 
This confinement is necessary to introduce spin correlations in the dot, such that only an electron with 
opposite spin can enter the TQD. The incoming current is then either $\uparrow$- or $\downarrow$-polarized,
depending on the position of the energy levels in the adjacent dot. 
The ac field frequency $\omega$ is in resonance with the central dot (dot 2), $\omega = \Delta_2$, in order
for the right dot (dot 3) to act as the filter dot. The TQD is here operated at the triple point $(2,0,0)\leftrightarrow(1,1,0)\leftrightarrow(1,0,1)$.
We restrict the discussion for simplicity to the case where the Zeeman splittings are $\Delta_1 = \Delta_3 > \Delta_2$,
although this condition is not necessary, as long as $\Delta_{1,3}\neq\Delta_2$.

From the previous sections we already know that depending on the
detuning, the dot connected to the drain can act as $\uparrow$- or
$\downarrow$-filter. In a TQD, there is one more degree of freedom compared to the DQD regarding the ``detuning'' between the dot levels.
Without loss of generality, we can fix the energy level of dot 1, and move the energy levels of
dot 2 and 3 (which is experimentally realized by applying gate voltages to the corresponding dots). 
Under these conditions, there are then four relevant energy level configurations, which are shown in Fig.~\ref{tqdinverter}, lower panel.
In two of the configurations (I and II), the TQD acts as a {\it spin-polarizer}, and in the other two (III and IV) the 
electron spin is {\it inverted}. 
We hereby arrive at another important result of our work: A TQD can be tuned as both spin-polarizer and
spin-inverter, by confining one electron in the left dot and adjusting the gate voltages at two of the three dots. 
Then electrons coming from the left lead can only enter with a distinct spin-polarization, which depends on the 
level position of the central dot. 
As the magnetic field $B_{_\text{ac}}$ is turned on with frequency $\omega = \Delta_2$, the electron
spin coming from dot 1 is rotated in dot 2, whereas dot 1 and dot 3 due to their different Zeeman splittings
are far off resonance from the ac field. Dot 3 then acts as spin-filter and, 
depending on the relative position of its energy levels with respect to dot 2, 
a ${\uparrow}$- or ${\downarrow}$-polarized current is produced, similar as in the DQD described in the previous sections. 

We plot the total $I_{\text{total}}$ and spin-resolved currents $I_{\uparrow}$ and $I_{\downarrow}$ versus
the two gate voltages applied to dot 2 and dot 3 in Fig.~\ref{tqdinverter}, together with sketches of
the corresponding energy level distribution. In situations I and IV, dot 2 is energetically in resonance with the $\uparrow$-level
in dot 1. Therefore, only $\uparrow$-electrons coming from the left lead will be able to tunnel to dot 2. Here they are inverted
due to $\omega = \Delta_2$, where the renormalized energy levels have been depicted schematically as we did for the DQD. 
It depends then on the level position of dot 3, if the outgoing current is spin-up (case I) or spin-down (case IV) polarized.
An analogue situation occurs for cases II and III: the energy level of dot 2 is such that only $\downarrow$-electrons can tunnel
from dot 1 to dot 2. Again, after rotation due to the ac field in dot 2, in dot 3 the spin is filtered without inversion (case II) or 
inverted (case III). 

\section{Conclusions}\label{conclusions}
In summary, we have analyzed spin current
polarization in the transport through a DQD with one extra electron, and through
a TQD with two extra electrons in the system.
The quantum dot arrays are subjected to two different external magnetic fields: 
an inhomogeneous dc field, which produces different Zeeman splittings in the dot, and a time dependent ac field, 
that rotates the electron spin in one dot, when the resonance condition $\omega = \Delta_{\text{Z}}$ is fulfilled.
For the DQD, we have analyzed both off-resonance and resonance conditions of the ac field with either
one of the Zeeman splittings. 
Our results show that ac magnetic fields produce strongly spin-polarized
current through a DQD depending on the detuning
of the energy levels in the dots and on the resonance conditions.

Finally, we have proposed a TQD in series as both $\emph{spin-polarizer}$
and $\emph{spin-inverter}$. 
As in a DQD, in a TQD different Zeeman splittings in the sample combined with a resonant ac 
frequency give way to spin-polarized currents. In addition, spin-polarized incoming
current can be achieved, and thus the spin-polarizing mechanism can be extended to a 
spin-inversion mechanism. 
Our results show that dc and ac magnetic fields combined with gate
voltages allow one to manipulate the current spin-polarization
through DQDs and TQDs which are then able to work as a spin-filter
and spin-inverter.

In spintronic devices at the nanometer scale an environment of 
nuclei introduces additional spin-flip processes
that can lower the efficiency of the desired mechanism.
In our set-up, we do not expect spin-flip processes due to hyperfine interaction to 
influence drastically on the results, because hyperfine spin-flip times are
usually much longer than typical tunneling times in quantum dot arrays, 
especially in finite magnetic fields, where the hyperfine interaction is an inelastic process.

Therefore, the systems presented in this work are promising candidates for spintronic devices.

\section*{Acknowledgements}
We are grateful to R. S\'{a}nchez, C. Creffield, J. Sabio and S. Kohler for 
helpful discussions and critical reading of the manuscript. We acknowledge financial support
through grant MAT2008-02626 (MEC), from JAE (CSIC)(M.B.) and from
ITN no. 234970 (EU).

\end{document}